\documentclass[notitlepage,superscriptaddress,nofootinbib,tightenlines,preprintnumbers,twocolumn]{revtex4-2} 
\usepackage{amsmath,verbatim,latexsym,amssymb,indentfirst,mathrsfs,mathtools,amsthm,bbm,bm,hyperref,url,cancel,subcaption,enumitem}
\usepackage[font=small,labelfont=bf,format=plain,justification=raggedright,singlelinecheck=false]{caption}
\usepackage{verbatim,indentfirst}
\usepackage[title,titletoc]{appendix}
\usepackage{array} 
\usepackage{graphicx}
\usepackage{graphicx,wrapfig,lipsum} 
\usepackage{afterpage}
\usepackage{epstopdf}
\newcommand{\wordcount}[1]{%
  \immediate\write18{texcount -total -merge -sum -q #1.tex  > #1.wcdetail }%
  \verbatiminput{#1.wcdetail}%
}

\renewcommand{\thesection}{\Roman{section}}
\renewcommand{\thesubsection}{\Roman{section} \Alph{subsection}}

\makeatletter
\def\p@subsection{}
\makeatother

\makeatletter
\def\p@subsubsection{}
\makeatother

\makeatletter
\newcommand\footnoteref[1]{\protected@xdef\@thefnmark{\ref{#1}}\@footnotemark}
\makeatother

\usepackage{soul}

\usepackage{float}
\usepackage{color}
\usepackage[dvipsnames]{xcolor}
\usepackage[normalem]{ulem}

\newcommand*{\changes}[1]{\textcolor{black}{#1}}
\newcommand*{\second}[1]{\textcolor{black}{#1}}

\begin{document}


\title{Dynamical excitation control and multimode emission of an atom-photon bound state}

\author{Claudia Castillo-Moreno}
\email{claudiac@chalmers.se}
\affiliation{Department of Microtechnology and Nanoscience, Chalmers University of Technology, 412 96 Gothenburg, Sweden}
\author{Kazi Rafsanjani Amin}
\affiliation{Department of Microtechnology and Nanoscience, Chalmers University of Technology, 412 96 Gothenburg, Sweden}
\author{Ingrid Strandberg}
\affiliation{Department of Microtechnology and Nanoscience, Chalmers University of Technology, 412 96 Gothenburg, Sweden}
\author{Mikael Kervinen}
\altaffiliation{Present address: VTT Technical Research Centre of Finland Ltd. Tietotie 3, Espoo 02150, Finland}
\affiliation{Department of Microtechnology and Nanoscience, Chalmers University of Technology, 412 96 Gothenburg, Sweden}
\author{Amr Osman}
\affiliation{Department of Microtechnology and Nanoscience, Chalmers University of Technology, 412 96 Gothenburg, Sweden}
\author{Simone Gasparinetti}
\email{simoneg@chalmers.se}
\affiliation{Department of Microtechnology and Nanoscience, Chalmers University of Technology, 412 96 Gothenburg, Sweden}


\begin{abstract}

Atom-photon bound states arise from the coupling of quantum emitters to the band-edge of dispersion-engineered waveguides. Thanks to their tunable-range interactions, they are promising building blocks for quantum simulators. Here, we study the dynamics of an atom-photon bound state emerging from coupling a frequency-tunable quantum emitter -- a transmon-type superconducting circuit -- to the band-edge of a microwave metamaterial. Employing precise temporal control over the frequency detuning of the emitter from the band-edge, we examine the transition from adiabatic to non-adiabatic behavior in the formation of the bound state and its melting into the propagating modes of the metamaterial. Moreover, we experimentally observe multi-mode emission from the bound state, triggered by a fast change of the emitter's frequency. Our study offers insight into the dynamic preparation of APBS and provides a method to characterize their photonic content, with implications in quantum optics and quantum simulation.

\end{abstract}

{\let\newpage\relax\maketitle}


Coupling quantum emitters to photonic lattices or metamaterials strongly modifies their spontaneous emission. When the emitter's frequency 
is close to a band-edge and within a band gap of the lattice, an atom-photon bound state (APBS) is formed -- a stationary excitation whose photonic component is exponentially localized around the physical location of the emitter. Because the localization length is controlled by the frequency detuning from the band-edge, APBS can mediate long-distance interactions with tunable range~\cite{douglas2015}. Following theoretical studies~\cite{douglas2015, gonzalez-tudela_subwavelength_2015, biella2015, 
lu2016a,calajo2016,martinezalvarez2019,belyansky2021a, fedorov2021b, ciccarello_resonant_2011, sheremet2023, fernandez-fernandez2022, romn-roche2020, calajo2019}, APBS have been observed in ultra-cold atoms coupled to photonic waveguides, optical lattices, and superconducting circuits~\cite{ hung2016, chang_colloquium_2018, 
bruzewicz_trapped_ion_2019, mcdonald2018, sundaresan2019, manovitz_quantum_2020, joshi_quantum_2020, carusotto2020, brehm2021, mirhosseini2018, vrajitoarea2024} and their properties have been leveraged to simulate spin models, prepare many-body correlated states, and explore many-body quantum phase transitions~\cite{morvan2022, joshi_quantum_2020, bello_spin_2022, zhang2023a,tabares2023, mcbroom-carroll2023, ferreira2024}.

Despite these advances, a dynamical characterization of individual atom-photon-bound states is still lacking. The (static) exponential localization of the photonic component has been characterized through their interaction with the metamaterial edges or among APBS clouds~\cite{vaidya2018, liu2016c, sundaresan2019}, or by coupling 
emitters to each resonator site~\cite{kim2021a, zhang2023a}. Additionally, time-dependent studies have 
observed non-Markovian dynamics~\cite{ferreira2021}, population exchanges~\cite{scigliuzzo2022}, and photonic hoppings through the metamaterial~\cite{ zhang2023a}.
These measurements involve fast ``quenches'' in which the photonic fraction of the APBS, as well as its localization length, are rapidly changed by varying the frequency of the emitter. Yet, the detailed dynamics, characteristic timing, and mode decomposition of these states remain unexamined. 

In this Letter, we study the time-dependent formation and melting of an APBS in a superconducting circuit. We combine dispersive measurements of the atomic population with frequency-resolved measurements of the radiation emitted by the APBS as the emitter is quenched. \changes{After the quench, the photonic part of the APBS, which was localized when the APBS was formed, turns into delocalized states in the metamaterial which then propagate and are measured at the output of our metamaterial. We refer to this process as “melting”.} 
We observe a crossover from adiabatic to non-adiabatic dynamics due to multi-state Landau-Zener tunneling to the frequency modes at the band-edge of the photonic band. Moreover, we characterize the emitted radiation when quenching the APBS, and detect multi-mode emission from up to 9 modes of the 
metamaterial. We find that an effective model captures overall trends, but precise dynamics and emission's spectral content are very sensitive to disorder in the metamaterial.
Our methodology is generally applicable to localized excitations of emitters coupled to photonic lattices and can facilitate the design of quantum simulators~\cite{douglas2015,zhang2023a,tabares2023} and topological interconnects~\cite{vega2023}.

\begin{figure}
    \centering
    \includegraphics[width=0.95\linewidth]{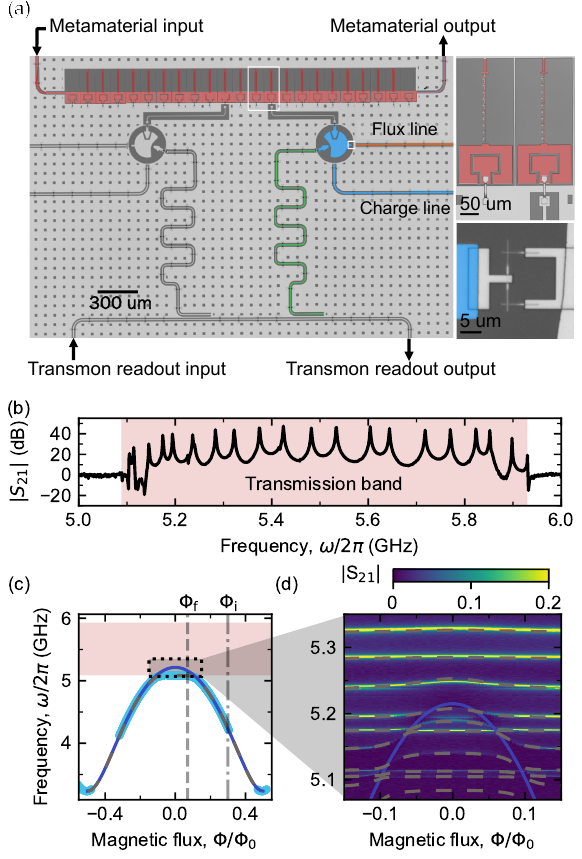}
    \caption{\textbf{Quantum emitter coupled to a metamaterial. (a)} False-color micrograph of the device. 
    Salmon: lumped-element resonator array forming the
    metamaterial. 
    Blue: transmon. 
    Green: readout resonator. Insets, in white, show (top) two resonators including the Josephson junction arrays, 
    and (bottom) 
    transmon's SQUID.
    \textbf{(b)} $\vert S_{21}\vert$ vs frequency. 
    Transmission band 
    in salmon. \textbf{(c)} Emitter frequency vs magnetic flux. Light blue: APBS frequency. 
    Dark blue: bare emitter frequency. Gray, dashed: APBS frequency from the effective model. Vertical gridlines \changes{illustrative} working points. \textbf{(d)} $\vert S_{21}\vert$ vs flux and frequency, with the response of the 9 first metamaterial modes to a change in the bare emitter frequency (dark blue). Dashed lines: model.} 
    \label{fig1}
\end{figure}

Our superconducting quantum circuit includes a metamaterial consisting of an array of 21 nearest-neighbor-coupled, lumped-element resonators~[Fig.~\ref{fig1}(a)]. Each resonator features an array of 10 Josephson junctions as the inductor, shunted by a capacitive element, resulting in a characteristic impedance $Z_r\approx390~\Omega$~\cite{scigliuzzo2022}. Input and output ports are capacitively coupled to the metamaterial's first and last sites to directly measure its transmission band and collect emitted radiation from the system. Two frequency-tunable, transmon-type artificial atoms~\cite{koch2007a, krantz2019} are capacitively coupled to the metamaterial resonators at sites 10 and 13. 
The two transmons are nominally identical and use asymmetric superconducting quantum interference devices (SQUIDs) as nonlinear inductors, resulting in two first-order flux insensitive points (sweet spots). The lower one resides well below 
the metamaterial's photonic band and the upper one inside it. 
Dedicated lines 
allow precise control of the transmon's frequency, via the flux (Z control), and population excitation, through the charge (XY control), while dispersively-coupled, frequency-multiplexed readout resonators allow measurement of the transmon populations~\cite{blais2021}. In the following, we use the transmon coupled to site 13~[false-colored in blue in Fig.~\ref{fig1}(a)] as a quantum emitter, while the other 
is kept at its lower sweet spot and does not participate in the presented experiments \changes{(Supplementary Information)}.

We characterize the metamaterial by measuring the transmission coefficient, $\vert S_{21}\vert$, 
while keeping the emitter far detuned from the transmission band, \changes{$\Phi/\Phi_0 = 0.5$}. This measurement reveals a transmission band between 5.088-5.93~GHz composed of 21 modes (detected as peaks), corresponding to the (hybridized) 21 resonators~[Fig.~\ref{fig1}(b)]. A tight-binding description of the system, assuming identical resonators of frequency $\omega_r$ and nearest-neighbor coupling $J$, predicts a transmission band in the frequency range $[\omega_r-2J,\omega_r+2J]$. Fitting this expression to the data, 
we extract $\omega_r/2\pi\approx5.5~\rm{GHz}$ and $J/2\pi \approx 211$~MHz. However, mode spacing and linewidths 
significantly deviate from the tight-binding prediction, a previously reported effect~\cite{scigliuzzo2022} that we ascribe to disorder in the resonators. 

To characterize the interaction of the emitter with the metamaterial, we sweep its bare frequency 
via the flux applied through the SQUID, $\Phi$, and measure its dressed frequency with two-tone spectroscopy
[Fig.~\ref{fig1}(c)]. Far from the photonic band, the data points follow the usual flux dependence of asymmetric 
transmons~\cite{hutchings_tunable_2017}. Near the band-edge, 
interactions with metamaterial modes cause deviations: 
the emitter frequency is prevented from entering the transmission band, signaling the formation of an APBS at the lower band-edge~\cite{calajo2016}~[Fig.~\ref{fig1}(c)], and, 
the frequency of several modes in the band gets shifted, 
interpreted as due to multimode strong-coupling in a finite-bandwidth waveguide~\cite{liu2016c}~[Fig.~\ref{fig1}(d)]. 

Metamaterial frequencies and their interaction with the emitter are 
sensitive to 
circuit disorder; 
therefore, an identical-resonators model cannot quantitatively reproduce them. 
To make contact with the spectroscopy data and model subsequent 
experiments, we use a model with the band's mode frequencies and their individual couplings to the emitter as free parameters. 

This effective model is described by the Hamiltonian
\begin{equation}
\begin{aligned}
H & =\sum_{n=1}^N \tilde{\omega}_n a_n^{\dagger} a_n + \omega_{q}(\Phi)\frac{\sigma_z}{2}   
+ \sum_{n=1}^N g_n\left(a_{n} \sigma_{+} + a_{n}^{\dagger} \sigma_{-}\right),
\label{eqn:model}
\end{aligned}
\end{equation}
in which $\tilde{\omega}_n$ are the dressed frequencies of the photonic modes, $a_n$ ($a_n^{\dagger}$) are their corresponding photon annihilation (creation) operators, $\omega_q(\Phi)$ is the flux-dependent emitter frequency, and $g_n$ the static couplings between the emitter and each photonic mode. We truncate the Hamiltonian to the single-excitation subspace 
(Supplementary Information). Additionally, we focus on 
the first $N=9$ modes from the lower band-edge, as the remaining ones do not show appreciable frequency shift when the emitter frequency is changed
~[Fig.~\ref{fig1}(d)]. 
Our model 
reproduces the spectroscopy data when using mode frequencies extracted from measurements with the emitter far detuned and best-fitted coupling strengths $g_n/2\pi = 3-25$~MHz.

\second{The emitter and the metamaterial interaction gives rise to an APBS, which consists of a superposition of an emitter-like and a photon-like excitation, mathematically defined as
\begin{equation}
    |\Psi_{\rm APBS}\rangle = c_E \sigma_+|g, 0\rangle + \sum_{i=1}^N c_i a_i^{\dagger}|g, 0\rangle 
    \label{APBS_equation}
\end{equation}}
\second{in which $c_E$ ($c_i$) is the amplitude coefficient of the emitter-like ($i$-th photon-like) state, $|g, 0\rangle$ is the ground state of the composite system, $\sigma_+$ the ladder operator, and $a_i^{\dagger}$ the creation operator \cite{calajo2016}. The probability coefficients depend on the detuning between the emitter and the band. Therefore, by manipulating $\omega_{q}$ through $\Phi$, we can study the formation and melting of the APBS between two points, an `emitter-like point', $\Phi_{\mathrm{i}}$, at which the emitter is away from the band-edge and the APBS responds as a two-level system, and a `photon-like point', $\Phi_{\mathrm{f}}$, at which the emitter 
is highly hybridized with the photonic band
~\cite{lombardo_photon_2014}~[Fig.~\ref{fig1}(c)].}

\begin{figure}
    \centering
    \includegraphics[width=0.95\linewidth]{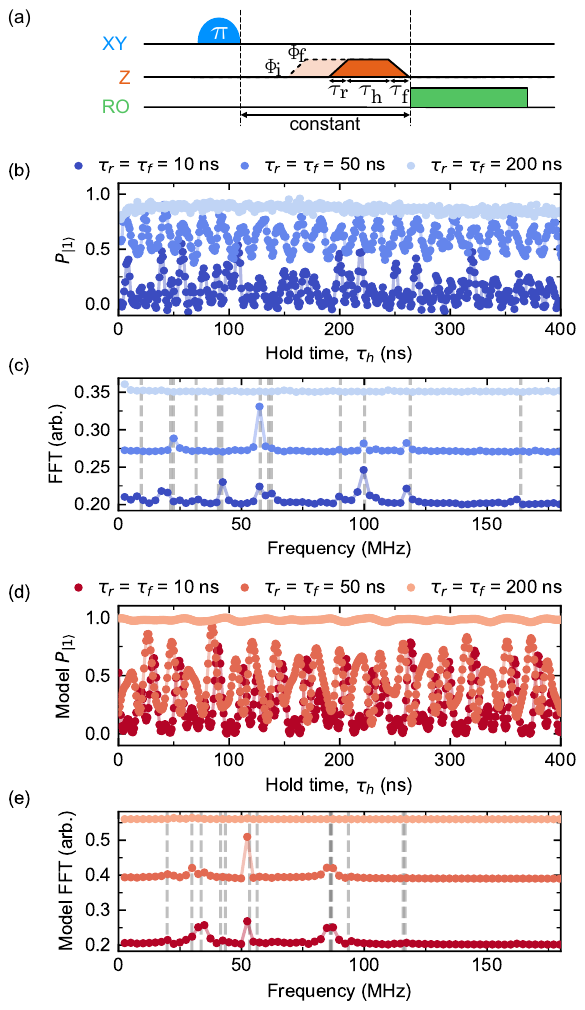}
    \caption{\textbf{APBS formation and melting at different speeds \changes{at $\omega_q(\Phi_{\mathrm{f}})\approx$~5.2~GHz}. (a)} Pulse sequence.
    \textbf{(b)} Transmon population vs hold time for $\tau_{\mathrm{r}}=\tau_{\mathrm{f}}=10,50,200$~ns. 
    \textbf{(c)} FFT of (b). Data shifted vertically for clarity. Dashed lines correspond to the frequency differences between the modes at 
    $\Phi_{\mathrm{f}}$. \textbf{(d,e)} Simulation results for 
    (b,c).
    }
    \label{fig2}
\end{figure}

APBS dynamics are explored by transitioning 
between these points at different speeds~[Fig.~\ref{fig2}(a)]. 
Starting at $\Phi_{\mathrm{i}}$, we excite the emitter with a $\pi$-pulse. Then, we apply a trapezoid-shaped flux pulse between $\Phi_{\mathrm{i}}$ and $\Phi_{\mathrm{f}},$
characterized by a rise time $\tau_{\mathrm{r}=10-200}$~ns, 
a hold time $\tau_{\rm hold}=0-400$~ns,
and a fall time $\tau_{\mathrm{f}=10-200}$~ns.
After returning to $\Phi_{\mathrm{i}}$, we measure the remaining population in the emitter, $P_{|1 \rangle}$, as a function of the time spent in $\Phi_{\mathrm{f}}$ for varying $\tau_{\mathrm{r}}$ and $\tau_{\mathrm{f}}$~[Fig.~\ref{fig2}(b)]. 
\changes{To remove the effect of decoherence, we interleave each measurement with a reference taken without the flux pulse, keeping the time delay constant.} 

As 
$\tau_{\mathrm{r}}$ and $\tau_{\mathrm{f}}$ decrease, $P_{|1 \rangle}$ decreases on average and oscillates with an increasing number of frequency components, as confirmed by a fast-Fourier transform (FFT) of $P_{|1 \rangle}$
~[Fig.~\ref{fig2}(c)]. For $\tau_{\mathrm{r}}=\tau_{\mathrm{f}}=200$~ns, $P_{|1 \rangle}$ is largely restored without oscillation. 
For $\tau_{\mathrm{r}}=\tau_{\mathrm{f}}=50$~ns, 
average $P_{|1 \rangle}$ 
decreases, and reproducible oscillations appear, as evident from FFT. For even shorter $\tau_{\mathrm{r}}=\tau_{\mathrm{f}}=10$~ns, 
$P_{|1 \rangle}$ decreases further, and more frequency components appear in the 
oscillations.

To understand these trends, we solve the time-dependent Schr\"odinger equation of the Hamiltonian in Eq.~\ref{eqn:model}, 
incorporating the time-dependent $\omega_{q}$ 
from the applied flux pulse and flux-to-frequency transfer function. Our model qualitatively reproduces the observed behavior, 
with decreasing population and increasing oscillations as the formation and melting speeds increase~[Fig.~\ref{fig2}(d),(e)].

In both the measurement and the simulation, oscillation frequencies 
align with differences between the system's dressed modes~[Fig.~\ref{fig2}(c),(e)]. This correspondence suggests that \changes{the coupling between the APBS and the metamaterial modes enables a non-zero probability of population transitions from one mode to the other, by Landau-Zenner tunneling ~\cite{shytov_landau-zener_2004, wubs_gauging_2006, saito_dissipative_2007, stehli2023}.} Therefore, for short $\tau_{\mathrm{r}}$/$\tau_{\mathrm{f}}$, the emitter's excitation gets distributed over several 
modes, with beatings in the emitter population 
due to quantum interference. 

In contrast, for long $\tau_{\mathrm{r}}$/$\tau_{\mathrm{f}}$, the large $P_{|1 \rangle}$
and the lack of oscillations indicate the population 
is adiabatically transferred to 
the APBS. The 
threshold for adiabatic transfer 
depends on the mode frequencies and their coupling strength, and it is coarsely estimated from the single-mode Landau-Zener formula
, $P_{\mathrm{LZ}}=\exp(-2\pi\Gamma)$, with $\Gamma=g^2 \Delta t/\Delta E$, in which $\Delta E$ is the difference between the emitter's initial and final energies, 
$\Delta t$ the $\tau_{\mathrm{r}}$, and $g$ the coupling between the emitter and each metamaterial mode. 
Our estimates show the adiabatic limit is reached for $\Delta t=200-300$~ns, 
matching observations.  

However, the model and data show two quantitative differences. 
\second{While the model predicts $P_{|1 \rangle}=1$ for $\tau_{\mathrm{r}}=\tau_{\mathrm{f}}=200$~ns, the data shows $P_{|1 \rangle}=0.9$.
We attribute this difference to population transfer to 
a coherent two-level system with a resonant frequency 
between $\omega_q(\Phi_{\mathrm{i}})$ and $\omega_q(\Phi_{\mathrm{f}})$ 
 (Supplementary Information).} Additionally, FFT peak frequencies and intensities do not exactly match~[Fig.~\ref{fig2}(d),(e)], 
possibly from an incorrect model-parameter estimate from the spectroscopy data or the uncorrected flux-line transfer function 
~\cite{rol2020}.

\begin{figure}
    \centering
    \includegraphics[width=0.95\linewidth]{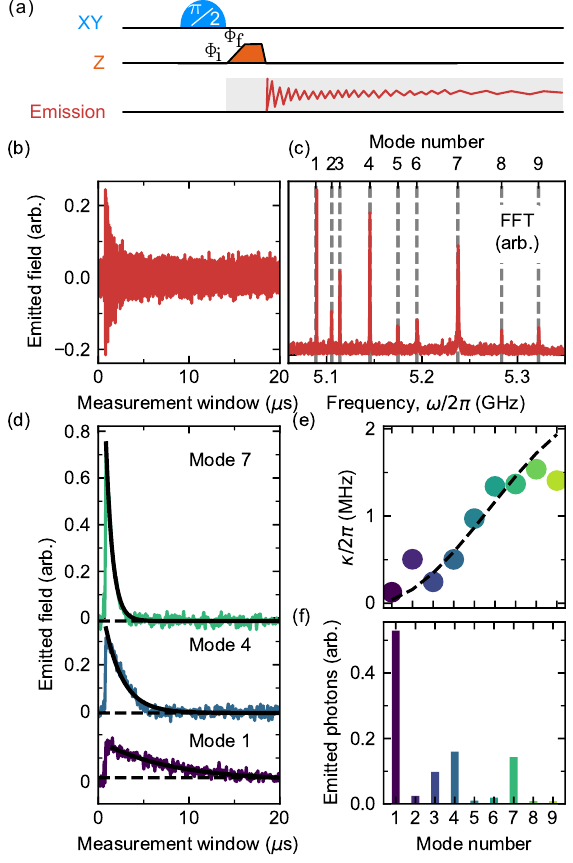}
    \caption{\textbf{Multimode emission \changes{at $\omega_q(\Phi_{\mathrm{f}})\approx$~5.2~GHz}. (a)} Pulse sequence.
    \textbf{(b)} Time trace of the emitted field. \textbf{(c)} FFT of (b). Vertical gridlines: metamaterial modes frequencies, extracted from Fig.~1(d). \textbf{(d)} Demodulated time traces for selected modes with exponential fits (black lines). \textbf{(e)} Filled circles: Emission decay rates vs mode index, extracted from fits in (d), dashed lines: tight-binding model predictions. \textbf{(f)} Emitted photons vs mode index}
    \label{fig3}
\end{figure}

The demonstrated dynamic control
enables direct access to the APBS's photonic component 
by melting the APBS following a 
quench. 
To do so, we adiabatically prepare the APBS 
by exciting the emitter with a $\pi/2$ pulse and then slowly ramping $\Phi$ between $\Phi_{\mathrm{i}}$ and $\Phi_{\mathrm{f}}$. \changes{$\tau_{\rm hold}=40~\rm{ns}$  for stabilization}. Then, we quickly ramp $\Phi$ back to $\Phi_{\mathrm{i}}$. At the same time, we record the coherent component, $\langle\hat{a}_{\rm out}\rangle$, of the outgoing field from the output port of the metamaterial, for 
$20~\mu$s~[Fig.~\ref{fig3}(a)]. 

The time trace~[Fig.~\ref{fig3}(b)] is digitally recorded with a 1-GHz-wide acquisition band centered at 5~GHz.
Its FFT reveals a total of 9 prominent peaks~[Fig.~\ref{fig3}(c)]. Notably, the frequencies of the peaks show a one-to-one correspondence with the frequencies of the 9 metamaterials's lowest-frequency modes~[Fig.~\ref{fig1}(b)]. 
We extract the temporal envelope of the radiation emitted into each mode by demodulating the time trace at each of the peak frequencies~[selected traces 
in Fig.~\ref{fig3}(d)]. The emission from each mode decays exponentially with a distinct decay rate~[Fig.~\ref{fig3}(e)] (Supplementary Information).
As a general trend, the decay rates are slower near the band-edge and become faster towards the center of the band.
\changes{This arises from states near the band-edge having a group velocity that approaches zero and, therefore, they interact less with the boundaries of the metamaterial~\cite{calajo2016, scigliuzzo2022} (Supplementary Information).} Additionally, a good agreement exists between our values and those predicted by the tight-binding model~[Fig.~\ref{fig3}(e)]. 
We quantify the 
emitted photons in each mode by integrating $\langle\hat{a}_n\rangle$ and squaring the result~[Fig.~\ref{fig3}(f)]. 

\changes{A 
quench in emitter frequency traps populations in the pre-quench instantaneous eigenstates. Subsequently, the photonic population is converted into propagating photonic modes in the metamaterial that decay to the output port.} Therefore, we interpret 
this measurement as a decomposition of the APBS's photonic part, providing a quantitative estimate of the relative probability densities in~Eq.\ref{APBS_equation}. Because the multimode emission stems from a single excitation in the APBS, we expect these modes to be entangled. However, a detailed study of the mode correlations is reserved for future research.

\begin{figure}[!h]
    \centering
    \includegraphics[width=0.95\linewidth]{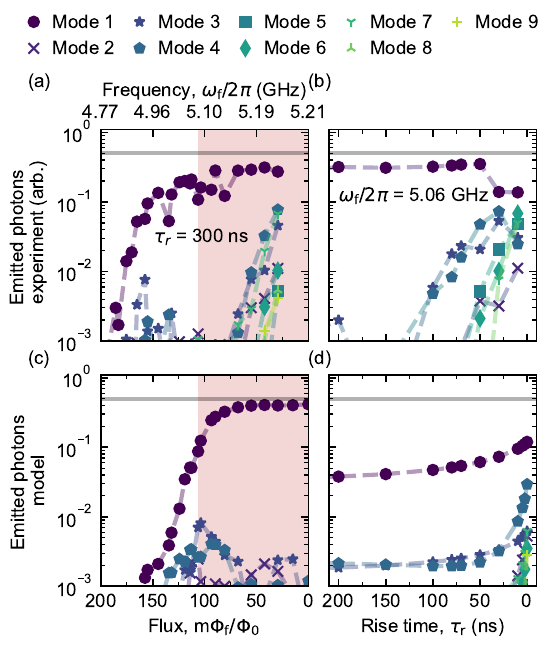}
    \caption{\textbf{Emission mode components depending on control parameters. \changes{Dashed lines connect data points for visual reference.} (a)} Emitted photons vs $\Phi_f$ for $\tau_{\mathrm{r}}=300~\rm{ns}$. Salmon: Transmission band. Data normalized so maximum emission equals 0.5 photons (horizontal gray line). \textbf{(b)} Emitted photons vs $\tau_{\mathrm{r}}$ \changes{at $\omega_f/2\pi= 5.06$~GHz} (inverted $x$-axis). \textbf{(c,d)} Numerical simulations for (a,b).
    }
    \label{fig4}
\end{figure}

We further explore the spectrally resolved emission of the APBS by varying the conditions for its formation. 
For different 
$\Phi_{\mathrm{f}}$ in the adiabatic regime, the emission mirrors the decomposition of the photonic component of the APBS. 
Far from the band, \second{the APBS is in the emitter-like state and the emission is weak. Once the frequency approaches the band-edge, the APBS becomes more photonic-like, and emission, mainly from the first mode, increases until it saturates. The contribution of higher-frequency modes depends on the specific realization of disorder in the array and varies non-monotonically when approaching the band-edge and rapidly increases once within the band with $\Phi_{\mathrm{f}}$ ~[Fig.~\ref{fig4}(a)].}


\second{The effect of $\tau_{\mathrm{r}}$ near the band-edge reveals a transition from single-mode to multi-mode emission with a sharp rise in the contribution of higher-frequency modes as $\tau_{\mathrm{r}}$ decreases~[Fig.~\ref{fig4}(a)]. 
We interpret this behavior as a direct excitation of propagating modes in the nonadiabatic regime.}


Our model captures the main observed trends~[Fig.~\ref{fig4}(c),(d)]. 
It reproduces the dominant emission from the lowest-frequency mode in the adiabatic regime~[Fig.~\ref{fig4}(c)] and the increased participation of higher-frequency modes at shorter $\tau_{\mathrm{r}}$~[Fig.~\ref{fig4}(d)]. However, there are quantitative differences. \second{In the adiabatic regime, close to the band there are small discrepancies in the emission participation of higher-frequency modes, and, the model does not reproduce the rapid 
emission rise within the band ~[Fig.~\ref{fig4}(a),(c)]. In the non-adiabatic regime~[Fig.~\ref{fig4}(b),(d)], the model predicts a shift from single-mode to multi-mode emission at shorter $\tau_{\mathrm{r}}$.} We ascribe these differences to a combination of factors.
First, the flux line's transfer function distorts flux pulses, 
\second{
affecting the speed at which the metamaterial modes are crossed and driving the system outside the adiabatic regime, which explains the sharp increase in participation of higher-frequency modes~[Fig.~\ref{fig4}(a),(b)].} 
Second, measuring undercoupled edge modes requires higher power, which can shift their frequencies due to the resonator’s nonlinearity, and affect the model parameters.
Lastly, disorder in the array may lead to uneven couplings to the ports and nonsymmetric spatial distribution of the propagating modes, affecting the relative strength of the detected emission \second{These two factors explain the discrepancies in the non-monotonic emission~[Fig.~\ref{fig4}(a),(c)] }\changes{(Supplementary Information)}. 

In conclusion, our study integrates emitter population measurements and frequency-resolved radiation detection
to elucidate the dynamic interaction between a quantum emitter and a metamaterial. The finite coupling of the emitter to the modes of the metamaterial results in a speed threshold 
for adiabatic 
excitation transfer from the bare emitter to the APBS. 
Understanding this threshold is important for using APBS in quantum simulators, 
as bound states have smaller anharmonicities than the bare emitters they originate from~\cite{sundaresan2019,scigliuzzo2022}, 
making adiabatic-state preparation 
potentially more advantageous than direct-pulsed excitation.
In addition, we have directly observed the melting of an APBS following a quench of the emitter's frequency, by detecting its emitted radiation. 
By resolving the frequency components of the emitted radiation, we gain direct access to the spectral decomposition of the APBS into its photonic components. This method can be applied to 
more exotic photonic lattices, multiple bound states coupled to the same lattice, or multi-photon bound states beyond the single-excitation subspace. 

We are grateful to Aamir Ali and Axel Eriksson for technical support, to Linus Andersson for fabricating the sample package, to Anita Fadavi Roudsari for her contributions to the nanofabrication recipes, to Alejandro Gonzalez-Tudela and Francesco Ciccarello for their input on the manuscript, and to Ariadna Soro and Anton Frisk Kockum for initial discussions. The Schr\"odinger equation was solved using QuTiP~\cite{johansson2013b}. The device in this work was fabricated in Myfab, Chalmers, a micro and nanofabrication laboratory. This work received support from the Swedish Research Council; and the Knut and Alice Wallenberg Foundation through the Wallenberg Center for Quantum Technology (WACQT). SG acknowledges financial support from the European Research Council via Grant No. 101041744 ESQuAT.


\nocite{ masluk2012, scigliuzzo2022, klimov_fluctuations_2018,burnett_decoherence_2019, lisenfeld_electric_2019, bilmes_resolving_2020, lisenfeld_electric_2019, bilmes_resolving_2020, rol2019, rol2020, jerger2019, rol2020, Sung2021}

\bibliography{Arrays}

\clearpage

\clearpage
\renewcommand{\thesection}{APPENDIX \Alph{section}}
\renewcommand{\thesubsection}{\Alph{section}.\arabic{subsection}}
\renewcommand{\thefigure}{S\arabic{figure}}
\renewcommand{\thetable}{S\Roman{table}}
\renewcommand\theequation{S\arabic{equation}}
\setcounter{figure}{0}
\setcounter{table}{0}
\setcounter{equation}{0}

\onecolumngrid

\appendix

\section*{Supplementary Information}




\section{Experimental setup}

\begin{figure}[h]
    \centering
    \includegraphics[width=0.8\textwidth]{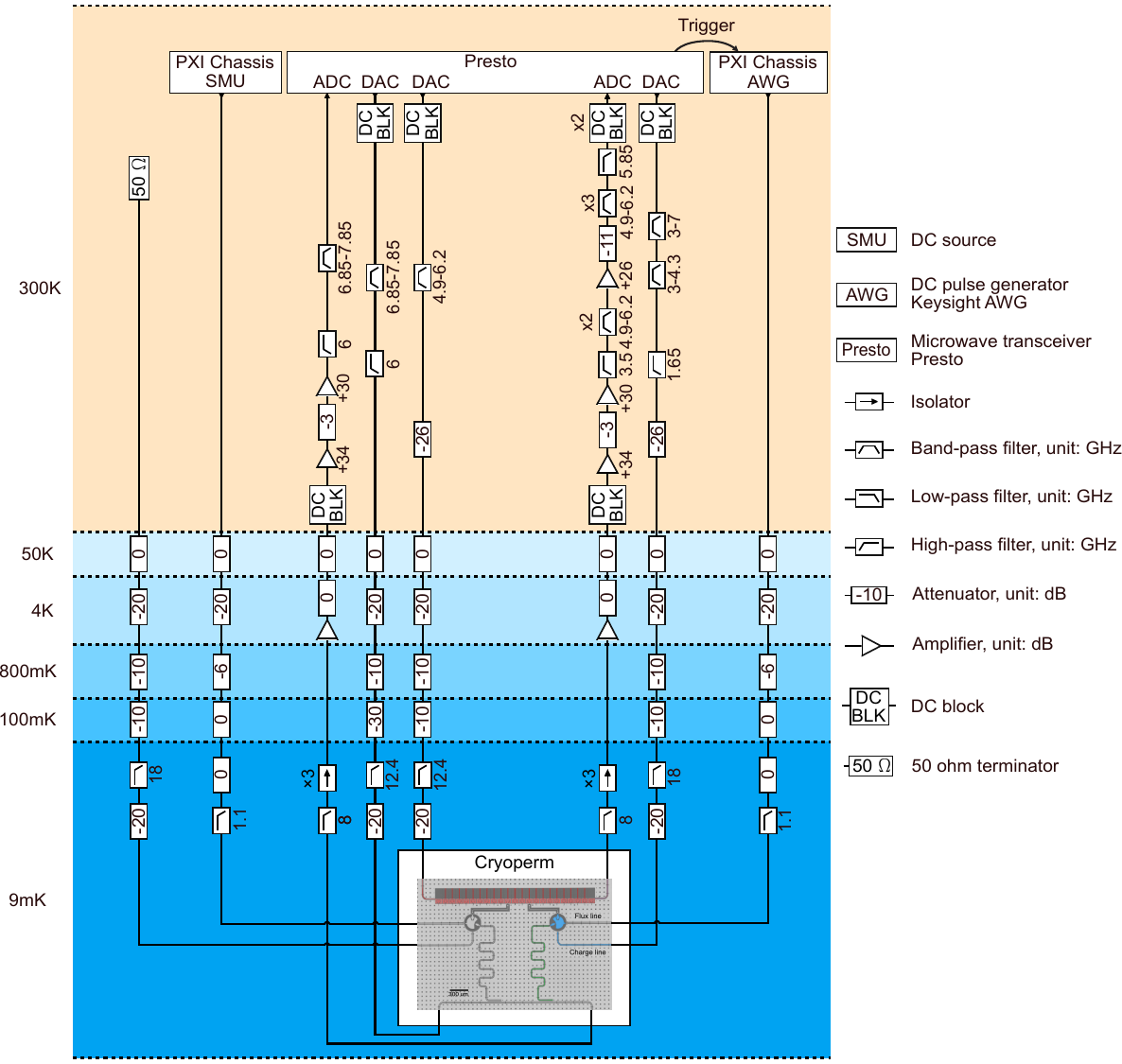}
    \caption{\textbf{Fridge connections.
    } \changes{Cooling-down stages are color-coded depending on the temperature, from yellow at room temperature to blue at the temperature in the mixing chamber. Each instrument is represented at the top and the lines connecting to the device, inside a cyoperm, include attenuators, amplifiers, and filters to improve the SNR.}}
    \label{fridge}

\end{figure}

\textbf{Supplementary information to "Dynamical excitation control and multimode emission of an atom-photon bound state"}

\vspace{1em}
Claudia Castillo-Moreno, Kazi Rafsanjani Amin, Ingrid Strandberg, Mikael Kervinen, Amr Osman, Simone Gasparinetti

\vspace{2em}

\changes{In~Fig.~\ref{fridge}, we present the full wiring diagram for our experiment, accompanied by a legend detailing the different instruments and microwave components.} The device under test is installed inside a dilution refrigerator that reaches a temperature below 9 mK. It is protected from electromagnetic interference with rf-tight copper shields, and from static magnetic fields with two Cryoperm shields and a superconducting shield. \changes{To minimize thermal photon generation, we attenuate incoming signals and use 0 dB attenuators as thermal anchoring points to dissipate heat along the coaxial cables. }The output lines feature multiple amplification stages, including a low-temperature High Electron Mobility Transistor (HEMT) and two (three) room-temperature amplifiers for the readout resonator (metamaterial). \changes{Additional filters and DC blocks are incorporated to mitigate noise outside our frequency range and prevent aliasing.}

\changes{The values below each element correspond to their specifications: amplification levels for the amplifiers given in dB, and the cutoff frequencies for the filters in GHz. Attenuator values are indicated within the elements themselves. To improve the cut-off response, we add filters in series with the number of each element included in series specified above the element.}
icrowave signals are sent and recorded using a microwave transceiver (Presto from Intermodulation Products AB) capable of direct digital synthesis in the band of interest. A Keysight PXI chassis is used to send both flux DC signals and flux DC pulses, through a Source Measure Unit (SMU) and an Arbitrary Wave Generator (AWG), respectively. We use the SMU DC signal in the flux line of the unused emitter to detune its frequency from the rest of the elements. In addition, we include a trigger from the Presto to the AWG to ensure that the different pulses are well synchronized between instruments.

{\color{black}

\section{Superconducting circuit sample design}

Our superconducting circuit sample, presented in~Fig.~1(a), includes a metamaterial formed of 21 high-impedance cavities. 
Each cavity includes an array of 10 Josephson junctions connected in series as inductor and a metal pad as capacitor. The high inductance of the Josephson junctions ensures high impedance and high coupling between cavities, which is crucial for the design of the metamaterial. The bandwidth of the transmission band is given by $4J$, with $J$ the coupling between cavities. 

The physical proximity controls the capacitive coupling between neighboring cavities. On the other hand, for the coupling between the metamaterial and the transmons we use a T-shaped metal pad, included in each element, and connect them through two airbridges to a long coupling element. This coupling design increases the distances between the components, reducing parasitic couplings. We add airbridges for the connection between the transmon and the resonator and the transmon and the metamaterial. These airbridges allow the ground metal plane to surround each element, avoiding parasitic currents, without reducing the couplings. Furthermore, we include shadow inner pads for each one of the cavities to maximize symmetry and reduce mismatches between the cavities. The two transmons are coupled to the physical center of the metamaterial, to sites 10 and 13. Therefore, their APBSs form in the center of the metamaterial and are isolated from the edges. For these measurements, we chose the transmon coupled to site 13 as it is closer to the output port, increasing the probability of collecting the emission. 

The coupling between the metamaterial modes and the transmon depends on the amplitude of the mode's spatial distribution at the emitter' physical coupling site. For instance, the emitter will be decoupled from a mode that has a node at the site of the emitter. In~Fig.~\ref{supplementary_couplings} we compare the theoretical couplings between the metamaterial modes and the two transmons in our system. The result shows that the distribution of the couplings is different for the two cases, as expected.

\begin{figure}[h]
    \centering
    \includegraphics[width=0.9\textwidth]{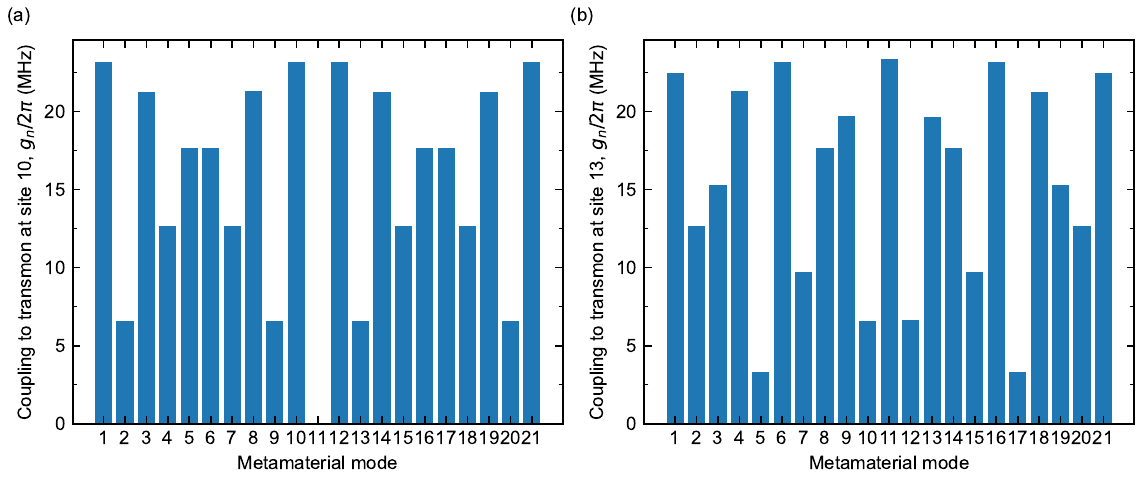}
    \caption{\textbf{Coupling distribution of each metamaterial to the transmon for two coupling sites. (a)} The transmon is coupled to the physical site 10 in the metamaterial, like the left transmon in our system [Fig.1 (a)]. \textbf{(b)} Transmon is coupled to site 13, like the right transmon in our system, the one under study [Fig.1 (a)]. }
    \label{supplementary_couplings}
\end{figure}

}
\section{Models and system parameters}

\subsection{Ideal tight-binding model}


The tight-binding model Hamiltonian assumes identical next-to-nearest-neighbor coupled resonators, and reads
\begin{equation}
\begin{aligned}
H & =\sum_{n=1}^N \omega_r a_n^{\dagger} a_n + \sum_{n=1}^{N-1} J \left(a_n^{\dagger} a_{n+1} + a_{n+1}^{\dagger} a_{n} \right) + J_{nn} \sum_{n=1}^{N-2} \left( a_{n+2}^{\dagger} a_{n} + H.c. \right) + \omega_{q}(\Phi)\frac{\sigma_z}{2}  +  g\left(a_{13} \sigma_{+} + a_{13}^{\dagger} \sigma_{-}\right),
\end{aligned}
\label{eq:tightbinding}
\end{equation}
Here, $\omega_r$ is the bare resonator frequency, $J$ and $J_{nn}$ are the nearest-neighbor and next to nearest-neighbor couplings between resonators, $a_n$ ($a_n^{\dagger}$) is the photon annihilation (creation) operator of the $n$-th resonator, $\omega_q(\Phi)$ is the flux-dependent transmon frequency, $\beta$ is the transmon's anharmonicity, $b$ and $b^{\dagger}$ its annihilation and creation operators, and $g$ the coupling between the transmon and the 13$^{\rm{th}}$ site of the metamaterial.

\subsection{Effective model}

Our effective model comes as an alternative to the tight-binding model to study the dynamics of our atom-photon bound state formation and melting. 
\changes{Theoretically, this effective model is derived by transforming the basis of the tight-binding model such that the metamaterial modes are diagonalized. 
However, instead of using the theoretical values from the tight-binding model, we directly include the system's disorder by replacing these values with the experimentally obtained frequencies of the resonators and their individual couplings to the transmon from transmission measurements. This approach addresses the disorder in the system, which causes variations in our metamaterial frequencies \cite{masluk2012, scigliuzzo2022}, rather than assuming identical frequencies for all resonators. The Hamiltonian for this model is given in~Eq.(1) of the main text. }

\subsection{Model parameters}

Table~\ref{characterization_table} lists the parameters extracted from the measurements using both the tight-binding and the effective models.
For the tight-binding model, the parameters $\omega_{r}$, $J$, $J_{nn}$ and $\kappa_r$ are obtained by comparing the model to the transmission measurement in our metamaterial in~Fig.~1(b). The coupling constant, $J$ is extracted from the bandwidth of the transmission band, $4J/2\pi = 842~\rm{MHz}$, and $\omega_{r}/2\pi$ corresponds to the center frequency. We corroborate these results by extracting both the coupling constant $ J/2\pi $ and the center frequency $ \omega_{r}/2\pi $ from microwave simulations of the system (To obtain the capacitances) and calibration measurements (For the inductances). In this case, $ \omega_{r}/2\pi \approx 5.37 $~GHz and $ J/2\pi\approx 190 $~MHz. These results show a good agreement with the ones extracted directly from our measurements. In addition, we obtain the coupling $J_{nn}$ from fitting the measured frequencies of the modes to those predicted by the model, and the decay rate, $\kappa_r$, from the linewidth of the center resonator.

We extract the impedance from  $Z_r = 1/\omega_r C_r$, with $C_r$ the simulated capacitance of the cavity. The obtained value is validated by comparing it to the expected impedance from the coupling constant, as $J\propto Z_r$. Finally, the coupling $g$ between the atom-photon bound state and metamaterial site 13 is obtained by fitting the experimental result of the atom-photon bound state spectroscopy~[Fig.~1(c)] to the frequency of the APBS from the Hamiltonian in Eq.\ref{eq:tightbinding}. 

For the case of the effective model, the individual frequencies of the metamaterial modes are obtained directly from the 21 peaks in the $\vert S_{21}\vert$ measurement in~Fig.~1(b). The couplings of the APBS to the metamaterial modes, $g_n$, require a more complex fitting. We use an optimizer that obtains the flux-dependent eigenstates from the Hamiltonian of the effective model and fits the results to the frequency of the atom-photon bound state in~Fig.~1(c) and measured flux-dependent metamaterial modes in~Fig.1(d). 

\begin{table}[h!]
    \centering
    \caption{Model parameters and experimentally determined values.}
    \begin{tabular}{lcc}
        \hline
         \textbf{Parameter} & \textbf{Symbol} & \textbf{Value} \\ \hline \hline
          \textbf{Qubit and resonator} \\
        Emitter frequency & $\omega_q/2\pi$ & 3.23-5.23 GHz \\
        Relaxation time at sweet spot & $T_1$ & 8.45 $\mu$s \\ 
        Coherence time at sweet spot & $T_2$ & 6 $\mu$s \\
        Anharmonicity & $\beta/2\pi$ & $-249.8~\rm{MHz}$ \\
        Readout resonator frequency & $\omega_{\rm{res}}/2\pi$ & 7.462 GHz \\
        Readout resonator decay & $\kappa_{\rm{res}}/2\pi$ & 178 kHz \\ \hline \hline
        \textbf{Tight-binding model} \\
        Metamaterial center frequency & $\omega_r/2\pi$ & 5.5075 GHz \\
        Nearest-neighbor coupling & $J/2\pi$ & 211.25 MHz \\ 
        Next-nearest-neighbor coupling & $J_{nn}/2\pi$ & 1.34 MHz \\ 
        Resonator decay & $\kappa_r/2\pi$ & 6.44 MHz  \\ 
        Resonator impedance & $Z_r$ & 387.23 $\Omega$\\
        Emitter-resonator coupling & $g/2\pi$ & 72.85 MHz \\
        \hline \hline
        \textbf{Effective model} \\
        Metamaterial frequencies & $\tilde \omega_n/2\pi$ & \{5.088,
        5.105,  
        5.114, 
        5.145, 
        5.174, 
        5.194, 
        5.236, 
        5.283, 
        5.322\}
        GHz \\
        Emitter-resonator couplings & $g_n/2\pi$ & \{ 20.67, 5.01, 9.33, 19.01, 3.06, 11.393, 25.434, 10.99, 20.16\} MHz \\ \hline
    \end{tabular}
     
\label{characterization_table} 
\end{table}

\section{Two-level-system spectroscopy}\label{supplementary_TLS}

Two-level systems (TLS) are intrinsic defects present in superconducting qubits \changes{that arise from material and interface defects. They degrade qubit performance by reducing its coherence, producing frequency fluctuations, and increasing relaxation rates}~\cite{klimov_fluctuations_2018,burnett_decoherence_2019, lisenfeld_electric_2019, bilmes_resolving_2020}. Therefore, to ensure reliable measurements, TLS need to be characterized and, if possible, avoided. 

\begin{figure}[h]
    \centering
    \includegraphics[width=0.6\textwidth]{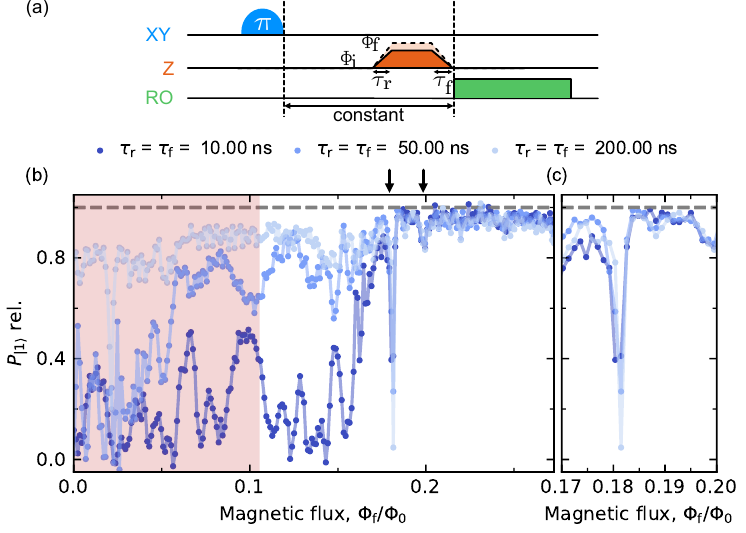}
    \caption{\changes{\textbf{TLS spectroscopy showing the loss of population depending on  $\tau_{\mathrm{r}}$ and $\tau_{\mathrm{f}}$.
    (a)} Pulse sequence: A $\pi$-pulse to the XY line, a trapezoidal-shaped flux pulse to the Z line with different amplitudes or $\Phi_{\mathrm{f}}$, and a readout pulse at a constant time from the $\pi$-pulse, to normalize decoherence in between the measurements. \textbf{(b)} Recovered population for three different rise and fall times. In shaded salmon the position of the metamaterial transmission band. Dashed line represents the full recovery. \textbf{(c)} Zoom-in of the two-level fluctuator position and effect in the recovered population.}}
    \label{TLS}
\end{figure}

In our case, we characterize the TLS landscape of the system by measuring the total population in the excited state, $P_{\vert1>}$ rel., at the initial frequency point, $\Phi_{\mathrm{i}}$, as the transmon frequency is swept with a flux pulse [Fig.~\ref{TLS}]. 
The pulse sequence is similar to the one in~Fig.~2(a). We excite the emitter with a $\pi$ pulse and apply a trapezoidal flux pulse with varying flux amplitudes, $\Phi_{\mathrm{f}}$, to sweep the final frequency. The transmon interacts with the TLS for 100 ns before being brought back to its initial frequency. 
Because the transmon coherence time, $T_1$, is reduced when coupled to a TLS, the recovered population will be reduced when a TLS is present. This measurement is similar to those conducted in~\cite{lisenfeld_electric_2019, bilmes_resolving_2020}.

\changes{We validate the results obtained in~Fig.~2 by repeating the TLS measurement for the three cases $\tau_{\mathrm{r}}=\tau_{\mathrm{f}}=$10, 50, 200 ns.} In all the cases, a visible reduction in the retrieved population happens at approximately 0.2~$\phi_0$ and 0.18~$\phi_0$, which we identify as two TLS, \changes{marked with the two arrows}. For the TLS at 0.2~$\phi_0$, its effect is small and all the population is recovered after crossing it for the three cases. However, in the case of the TLS at 0.18~$\phi_0$, we obtain an average population reduction of 10 $\%$. We attribute to this second TLS the loss in the recovered population when the atom-photon bound state is adiabatically created and melted in the main text~[Fig.~2]. 


\section{Spectroscopy result with tight-binding and effective model}

\begin{figure}[h!]
    \centering
    \includegraphics[width=\textwidth]{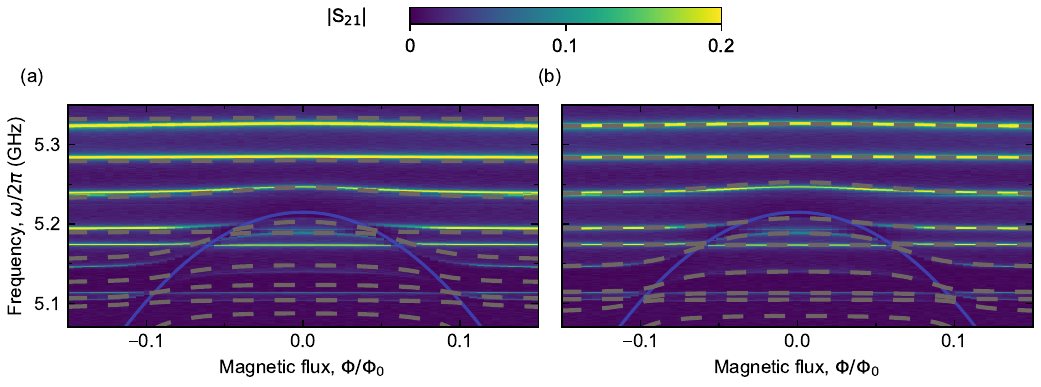}
    \caption{\textbf{Model results compared to transmission band experiment depending on applied magnetic flux on the transmon. (a)} 
    Tight-binding model eigenmodes (gray dashed lines). Discrepancies are visible for the first four modes from the bottom. \textbf{(b)} Effective model eigenmodes (gray dashed lines). Less visible discrepancies than for the tight-binding model. 
    }
    \label{supplementry_fluxmodels}
\end{figure}

\changes{To verify the selection of the effective model,} we compare the results of our two models to the measured metamaterial spectroscopy as a function of the applied magnetic flux to the transmon~[Fig.~\ref{supplementry_fluxmodels}(a,b)]. 

Disorder in the Josephson junctions affects the distribution of the frequencies in our metamaterial, which cannot be captured by the ideal tight-binding model.~[Fig.~\ref{supplementry_fluxmodels}(a)]. \changes{The deviation between the tight-binding model and the measured modes is larger for the modes at the band edges than those closer to the center of the band. This effect is especially visible in the four lowest frequency modes, which do not overlap with the real position of any of the modes.}

\changes{In contrast, in the case of our effective model, which is developed to account for the disorder at the price of a large number of free parameters, the overlap between the measured and theoretical data is evident and the model reproduces} the data more accurately~[Fig.~\ref{supplementry_fluxmodels}(b)].


\section{Exponential decay of the emitted field} 

After the quench and melting of the APBS, the electromagnetic field is detected at the output port, and the signal is demodulated to identify the emission from each photonic mode [Fig.~3~in the main text]. The outcoming field exhibits an exponential decay with a lifetime, $\tau$, which is related to the linewidth of the metamaterial modes, by $\kappa_n = 2\pi/\tau_n$.

\begin{figure}[h!]
    \centering
    \includegraphics[width=0.8\textwidth]{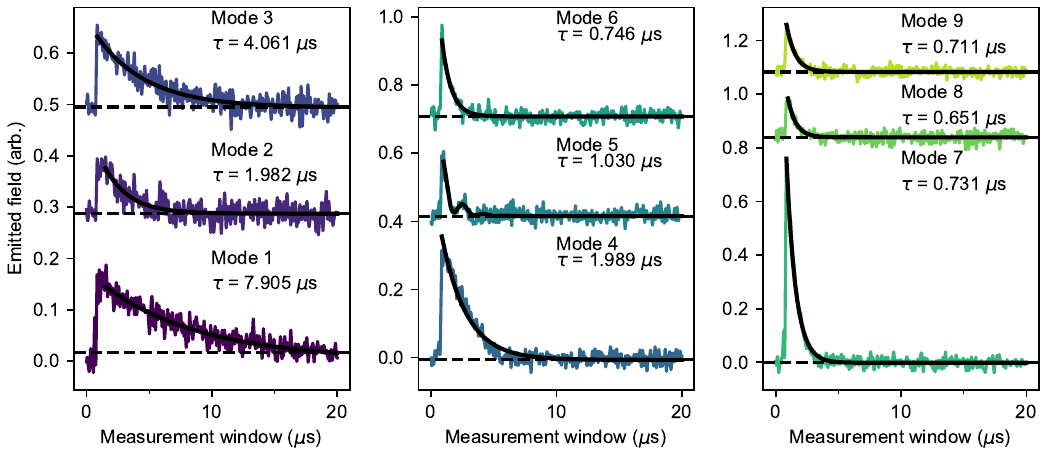}
    \caption{\textbf{Demodulated signal decays with their corresponding decay constant.} \changes{The signals have been shifted in the $y$-axis for comparison}. The dashed lines represent the baseline of the emission for each signal, while the colored traces depict the resulting demodulated emission from the different modes. Time step is 1 ns. The fittings are indicated in black, with $\tau$ representing the decay constant.
    }
    \label{supplementary_decay}
\end{figure}

\changes{To extract the value of $\tau$ for each mode, we fit the demodulated signals with an exponential decay in~Fig.~\ref{supplementary_decay}. The obtained values reveal a consistent trend: The modes at the band-edge show a maximal lifetime and, higher mode numbers correspond to smaller lifetimes. This trend is expected as the modes at the band-edge are longer lived than those in the center of the band because of vanishing group velocity as the dispersion relation curvature flattens out. A low group velocity means that the energy of the mode is more localized, which implies a reduction of the interaction between the mode and any imperfection or boundaries.  We attribute the small deviation from this trend in modes 2, 5, and 8 to a smaller signal-to-noise ratio due to their smaller participation in the emission.}

In addition, for the fifth mode, the emitted signal displays not only exponential decay but also a revival. Although we leave the characterization of this effect to further studies, we hypothesize that this revival may arise from a less-coupled mode to the ports, which causes revivals in its population triggered by the reflection of the signal at the ports.

{\color{black}

\section{Discrepancies between measurements and theory} 

Our experimental device comprises 25 modes in the 4-8 GHz band: the 21 coupled resonators, the two qubits, and their readout resonators. The bare resonant frequencies of the resonators forming the metamaterial are designed to be nominally identical; in practice, however, they are affected by disorder from the fabrication process of the Josephson junctions forming the inductive part of the resonators, and possibly by coupling to spurious modes on the chip and package. This disorder affects both the spatial distribution and the frequency distribution of the metamaterial modes, to which the experiments presented in this work are highly sensitive. As a result, in contrast to previous work, an effective description of the system (assuming, for example, identical frequencies and constant coupling rates for the resonators in the metamaterial) is insufficient to explain our observations; instead, we need to include a specific realization of the disorder in the model. Even when we do that, we find only a qualitative agreement between the data and the model. Below we suggest possible reasons for this discrepancy and suggest how they could be addressed in future work.


\subsection{Extraction of coupling parameters}

The coupling parameters in our model in Eq.~(1) are directly extracted from the spectroscopy measurement 
in~Fig.1(d). However, the signal from some of the edge modes is weaker than in the center of the band, due to the fact that these modes are more localized. To detect these modes, we drive them at higher power, which can cause frequency shifts due to the nonlinearity of the metamaterial modes. Such frequency shifts could lead to an erroneous estimate of the coupling parameters.
In future studies, this problem could be mitigated by using an amplification chain with better signal-to-noise ratio, for example one including a near-quantum limited parametric amplifier, or by designing resonator metamaterials with lower nonlinearity.

\subsection{Uneven coupling to the measurement ports}

In our system, the presence of disorder influences how the modes are spatially distributed, resulting in some modes having stronger coupling to one port than the other.
Because of this uneven coupling, the fraction of emitted radiation which is collected for each mode will depend on its relative coupling to the monitored port. As a result, the emission data presented in Fig.~3 and Fig.~4 may suffer from systematic errors.
To address this issue, we suggest monitoring both ports.

\subsection{Distortion in the flux lines}

Flux pulses experience distortions due to the various components in the flux line. A dominant distortion arises from low-pass filters causing square pulses to be transformed into a more sinusoidal shape, thereby altering their effect on the device \cite{rol2019, rol2020}.
To mitigate the effects of distortions, we suggest calibrating the flux line and applying a compensation (predistortion) to the flux pulses, as suggested in the literature~\cite{jerger2019, rol2020, Sung2021}.



}


\clearpage
\twocolumngrid

\end{document}